\documentclass[article]{aa} 

\usepackage[latin1]{inputenc}     
\usepackage[english]{babel}       
\usepackage{amssymb}              
\usepackage{amsmath}
\usepackage{amsfonts}             
\usepackage{graphicx}             
\usepackage{epstopdf}             
\usepackage{subfigure}
\usepackage{multirow}
\usepackage{array}                

\usepackage{natbib}
\usepackage{epsfig}

\def\ut#1{\mathop{\vtop{\ialign{##\crcr
     $\hfil\displaystyle{#1}\hfil$\crcr\noalign
     {\kern1pt\nointerlineskip}\hbox{$\hfil\sim\hfil$}\crcr
     \noalign{\kern1pt}}}}}

\def\undersymbol#1#2{\mathop{\vtop{\ialign{##\crcr
     $\hfil\displaystyle{#2}\hfil$\crcr\noalign
     {\kern1pt\nointerlineskip}\hbox{$\hfil#1\hfil$}\crcr
     \noalign{\kern1pt}}}}}
\def\arcsec{^{\prime\prime}}
\def\arcmin{^{\prime}}

\newcommand{\Wmap}{{\sc Wmap}}

\newcommand{\Boomerang}{{\sc BOOMERanG}}

\begin{document}

  \title{Measurement of the Crab nebula polarization at 90 GHz as a calibrator for CMB experiments}

      \author{J.~Aumont~\inst{1,8} \and 
	L.~Conversi~\inst{2} \and
	C.~Thum~\inst{3} \and	
	H.~Wiesemeyer~\inst{4}\and
	E.~Falgarone~\inst{5} \and	
	J.~F.~Mac\'{\i}as-P\'erez ~\inst{6} \and
	F. Piacentini~\inst{7} \and 
	E.~Pointecouteau~\inst{8} \and
	N.~Ponthieu~\inst{1} \and	
	J.~L.~Puget~\inst{1} \and	
	C.~Rosset~\inst{9} \and
	J.~A.~Tauber~\inst{10} \and	
	M.~Tristram~\inst{11}}	
              \institute{Institut d'Astrophysique Spatiale, Centre
Universitaire d'Orsay, Bât. 121, 91405 Orsay Cedex, France 
\and 
European Space Astronomy Center, P.O.Box 78, 28691 Villanueva de la
Ca\~{n}ada (Madrid), Spain 
\and
IRAM - Institut de Radioastronomie Millimétrique, 300, rue de la
Piscine, 38406 Saint-Martin d'Hères, France 
\and 
IRAM - Institut de Radioastronomie Millimétrique, Avenida Divina Pastora, 7, Núcleo Central, E-18012 Granada, Spain
\and 
LERMA/LRA, CNRS UMR 8112, École Normale Supérieure and Observatoire de
Paris, 24, rue Lhomond, 75231 Paris Cedex 05, France 
\and 
Laboratoire de Physique Subatomique et de Cosmologie, Université Joseph Fourier Grenoble 1, CNRS/IN2P3, Institut National Polytechnique de Grenoble, 53, avenue des Martyrs, 38026
Grenoble, France 
\and 
Dipartimento di Fisica, Università di Roma ``La Sapienza'', Rome,
Italy 
\and
Centre d'Étude Spatiale des Rayonnements, Université de Toulouse,
CNRS, 9, av. du Colonel Roche, BP44346, 
31028 Toulouse Cedex 4, France 
\and 
APC, Université Denis Diderot-Paris 7, CNRS/IN2P3, CEA, Observatoire
de Paris, 10, rue A. Domon \& L. Duquet, Paris, France 
\and European
Space Agency, Astrophysics Division, Keplerlaan 1, 2201AZ Noordwijk,
The Netherlands 
\and 
LAL - Laboratoire de l'Accélérateur Linéaire, Université Paris-Sud 11,
CNRS/IN2P3, Bât 200, 91898 Orsay Cedex, France}

   \date{Submitted: XXX; Accepted: XXX}

{
  \abstract
   {CMB experiments aiming at a precise measurement of the CMB
polarization, such as the Planck satellite, need a strong polarized absolute
calibrator on the sky to accurately set the detectors polarization
angle and the cross-polarization leakage. As the most
intense polarized source in the microwave sky at angular scales of
 few arcminutes, the Crab nebula will be used for
this purpose.} 
  {Our goal was to measure the Crab nebula polarization
characteristics at 90 GHz with unprecedented precision.}
   {The observations were carried out with the IRAM 30m telescope
employing the correlation polarimeter XPOL and using two orthogonally
polarized receivers.}
   {We processed the Stokes $I$, $Q$, and $U$ maps from our observations in
order to compute the polarization angle and linear polarization
fraction. The first is almost constant in the region of
maximum emission in polarization with a mean value of $\alpha_{\rm
Sky}=152.1\pm 0.3^\circ$ in equatorial coordinates, and the second is found to reach a maximum of $\Pi=30$\% for the most
polarized pixels. We find that a CMB experiment having a 5 arcmin
circular beam will see a mean polarization angle of $\alpha_{\rm
Sky}=149.9\pm0.2^\circ$ and a mean polarization fraction of $\Pi=8.8\pm0.2$\%.}
   {}
}

\authorrunning{Aumont et al.}
\offprints{J.~Aumont ({\tt jonathan.aumont@ias.u-psud.fr})}
\titlerunning{Crab nebula polarization at 90 GHz as a calibrator for CMB experiments}
   \keywords{ISM: supernova remnant, Polarization, (Cosmology:)
cosmic background radiation}

   \maketitle
%

\section{Introduction}

The Crab nebula (Tau A, M1 or NGC1952, at
coordinates $\alpha = 5^{\rm h} 34^{\rm m} 32^{\rm s}$ and $\delta =
22^\circ
0\arcmin 52\arcsec$, J2000) is a supernova remnant that
emits a highly polarized signal due to both the synchrotron emission of the
central pulsar and its interaction with the surrounding gas \citep[see
{\it eg}][]{2008ARA&A..46..127H}. In this
paper, we present the measurements of intensity and polarization of
the Crab nebula at 90~GHz (3.3 mm) performed at the IRAM 30~m telescope using
the XPOL instrument \citep{2008PASP..120..777T}. The strength of
this instrument is the precision in the angle of polarization with
respect to the sky reference frame, which highly improves
previous measurements. This precision is needed in particular to
use the Crab nebula as an absolute polarization calibrator for cosmic
microwave background (CMB) experiments, such as the Planck ESA mission. The
results for circular polarization are irrelevant for current CMB
polarimetry experiments and will be published elsewhere
\citep{IRAMCrabV}.

The last Thomson scattering of cosmological photons by free
electrons at the time of decoupling between matter and radiation at
$z_{\rm dec}=1088.2\pm 1.1$ \citep{2009ApJS..180..306D} have led to a
linearly polarized CMB radiation at a level of
a few percent. The linear polarization pattern on the sky can be
decomposed into a scalar and a pseudo-scalar field, respectively called $E$- and
$B$-modes, which have the advantage of being independent of
the reference frame
\citep{1997PhRvD..55.1830Z,2001PhRvD..64j3001Z}. This decomposition is
particularly useful because primordial density fluctuations (scalar
perturbations) can only produce $E$ polarization, while $B$ can only be
produced by lensing effects, exotic phenomena such as cosmological
birefringence \citep{1990PhRvD..41.1231C} and by the stochastic background of gravity
waves arising from the inflation epoch (tensor perturbations). The
detection of the latter would improve the
constraints on the inflationary model parameters in an invaluable
manner and constitutes one
of the most ambitious goals of observational cosmology.

The first measurement of the CMB $E$ polarization was
made by the 30 GHz radio interferometer, DASI, in 2002
\citep{2002Natur.420..772K}. Since then, in addition to a further measurement by
the DASI experiment \citep{2005ApJ...624...10L}, $E$ modes measurements have
been made with the CBI \citep{2004Sci...306..836R}, CAPMAP \citep{2005ApJ...619L.127B},
\Boomerang\ \citep{2006ApJ...647..813M}, \Wmap\ \citep{2007ApJS..170..335P}, {\sc MAXIPol}
\citep{2007ApJ...665...55W}, {\sc QUaD} \citep{2008ApJ...674...22A}
and {\sc Bicep} \citep{2009arXiv0906.1181C}. With these measurements,
the $C_\ell^{EE}$ angular power spectrum has now been detected over a
wide range of angular scales ($30\lesssim\ell\lesssim1000$). Additionally, the \Wmap\ satellite has given a
very precise measurement of the temperature-$E$ polarization
correlation, the $C_\ell^{TE}$ angular cross power spectrum, at large
scales up to multipoles $\ell\simeq500$ \citep{Nolta:2009zh}. To date, no detection of non-zero $C_\ell^{BB}$ angular power spectrum
has been reported and only upper limits on the effect of gravity waves from other spectra
can be inferred. The best indirect limit to date on the tensor to
scalar ratio $r$, given by \Wmap\ data and other
cosmological observables together is $r<0.2$ (95\% C.L.) \citep{2009ApJS..180..330K}.

Upcoming experiments target the detection of lower values of the
tensor-to-scalar ratio. The low signal of the $B$-modes compared to the
temperature and $E$-modes signals will require a high control of
systematics and good knowledge of the instrument. In particular, the
knowledge of its polarization characteristics will require a highly
and well known polarized source. 

The Crab nebula represents the
most suitable candidate as absolute calibrator for the polarization
angle and the linear polarization degree, given that it is the
most intense polarized source in the microwave sky at angular scales
of few arcminutes \citep{1979MNRAS.189..867F,1981MNRAS.194..961F,1983MNRAS.204.1285F}. 

The paper is organized as follows: in \S~\ref{calibration} we describe
the required calibration accuracy for a CMB experiment aiming at
the detection of the $B$-modes; in \S~\ref{instrument} we briefly describe
the XPOL instrument; \S~\ref{obs} reports the observations, data
reduction and systematic error analysis; in \S~\ref{results} we give our results; we discuss the
extension of our measurements to other frequencies in
\S~\ref{other_freq} and summarize our results in \S~\ref{conclusions}. 


\section{Required calibration accuracy}\label{calibration}

The expected primordial $B$-mode level at large angular scale, for $r=0.01$, is
typically 1\% of the $E$-mode level. More precisely, for $\ell < 20$, the
$E$-mode is on the order of $0.07\,\mu\mbox{K}^2$ \citep[from the WMAP
  best-fit model, see][]{Nolta:2009zh}, while the $B$-mode is expected
at $10^{-4}\,\mu\mbox{K}^2$. For comparison, the temperature
anisotropies are at the level $1000\,\mu\mbox{K}^2$ at large scale.
The Planck High Frequency Instrument, for example, has the potential
to detect or constrain the tensor-to-scalar ratio down to $r\simeq
0.03$ \citep{Efstathiou:2009kx}. However, such a low level will be
reachable only if the systematics are well controlled, otherwise,
given the hierarchy of signal levels, $T\gg E\gg B$, we may
expect leakage from temperature to polarization and a leakage of
$E$-mode to $B$-mode.

Various instrument parameters, if not precisely known, will induce
systematic effects. In the case of the Planck HFI polarization
sensitive bolometers, the detectors are sensitive to the power carried
by the electric field aligned with the detector's grid. The signal $s$
measured by a detector can be written \citep{Rosset:2009}

$$s = g\int \bigg[ I + \rho \big( Q \cos
2(\alpha+\alpha_d)\nonumber$$
\begin{equation} 
+U\sin
  2(\alpha+\alpha_d)\big) \bigg] A(\mathbf{n}) d\mathbf{n},
\end{equation}
where $I$, $Q$ and $U$ are the Stokes parameters of the sky signal
(intensity and linear polarization in the sky reference frame), $g$ is the detector gain, $A$ is
the instrument beam, $\rho$ is the detector polarization efficiency,
$\alpha$ is the angle between the telescope reference frame and the
local meridian in the sky, $\alpha_d$ is the angle at which the
polarimeter is maximally sensitive to polarized radiation, in the
telescope reference frame (here, we have made the simplification that the beam for
polarization is identical to the intensity beam, only modulated by the
sine and cosine of the detector orientation) and $\mathbf{n}$ is
a direction on the sky. In this simplified
model, $g$ and $A$ are measurable with unpolarized light (cosmological
and orbital dipole, planets). Only the
polarization specific parameters, $\alpha$ and $\rho$, require a known
polarized signal to be calibrated. Measuring with precision these
properties in laboratory or with an artifical source can be difficult
\citep[see {\it eg}][]{Rosset:2009,Takahashi:2009}, so measuring them
with a known astrophysical source allows cross-checking or even a
better calibration. In any case, the best limit on the precision of
the polarization efficiency and the orientation of detectors will be
set by the precision on the polarization fraction and polarization
orientation of the source.

In the case of a unique detector scanning the sky, it has been shown
\citep{ODea:2007kx,Rosset:2009} that the error on the $B$-mode
power spectrum due to an error $\Delta\alpha$ in the polarization
orientation and an error $\Delta\rho$ on the polarization efficiency is given by:
\begin{equation}
\Delta C^{BB}_\ell = 2\Delta\rho C^{BB}_\ell + 4\Delta\alpha^2 C^{EE}_\ell
\end{equation}
For example, an experiment targeting a tensor-to-scalar ratio
measurement of $r=0.01$, using the large angular scale $B$-mode (at
$\ell < 20$), will require the polarization orientation and the polarization efficiency
to be known with a precision of:
\begin{equation}
\Delta\alpha < 0.9^\circ,\qquad \Delta\rho<0.05,
\end{equation}
in order to have a leakage from $E$ to $B$-modes lower than one tenth of
the expected $C_\ell^{BB}$.

These tolerances set the precision needed for the calibrator, and
define our goal for the Crab nebula observations presented in this
paper.

\section{Instrumental setup}\label{instrument}

\subsection{General}

For the observation of the Crab nebula, we used the IRAM 30 m telescope at 89.189~GHz, the frequency of the
HCO$^+$(1--0) transition, where the angular resolution (the full width at half
power of the near--Gaussian beam) is 27''. We used the following IRAM receivers: A100 cryostat
(vertical linear polarization with respect to Nasmyth reference frame) and 
B100 cryostat (horizontal linear polarization). The mean system temperatures
were 87~K (A100) and 113~K (B100) while mean receiver temperatures
(single-sideband) were 67~K (A100) and 59~K (B100). The effective cold
loads used for temperature calibration had a mean temperature of 87~K
(A100) and 98~K (B100). We used the VESPA
backends in XPOL mode (see \S~\ref{subsec_xpol}),
with a 500~MHz bandwith (200 channels spaced by 2.5~MHz). Finally, the conversion factor from antenna temperature outside atmosphere to flux
density was 1~K $= 6.0$~Jy and the mean zenith opacity was $\tau_z$=0.06 at our observing frequency. 

\subsection{XPOL}
\label{subsec_xpol}

The signals from the two orthogonal
linearly polarized heterodyne receivers were detected in auto- and
cross-correlation, from which the four Stokes parameters were derived as
described in detail by \cite{2008PASP..120..777T}. This procedure, designated
XPOL, employs a precise calibration of the phase between the two receivers,
resulting in a phase error $<1^\circ$ per channel of 2.5 MHz, across the 500 MHz bandwidth.
 
Particular care was taken in the absolute calibration of the
polarization angle $\tau$. This angle was re-measured, for the purpose
of this experiment in the Nasmyth cabin, with respect to its horizontal
axis, where the receivers are stationary, as

\begin{equation}
 \tau = \frac{1}{2} \arctan (\frac{U}{Q}),\label{def_angle}
\end{equation}

where $Q$ and $U$ refer to the quantities measured in the Nasmyth reference frame. The precision with which the receivers, labelled H and V, actually
measure horizontally and vertically polarized power is determined by
the orientation of the grid (G3 in Fig. 1 of \cite{2008PASP..120..777T}) which
splits the incoming beam into a transmitted fraction which is
horizontally polarized, and a reflected fraction which is vertically
polarized. The horizontality of the incoming beam, and thus also that of the
H-fraction rests on the correct alignment of the telescope Nasmyth
mirror M3, the subreflector, and the validity of the implemented homology
corrections and the pointing model. Their combined error is less than 1 arcmin. The verticality of the V-fraction additionally depends on
the orientation of G3 whose normal should  make an angle of 45$^\circ$ with
respect to the local vertical. The mounting frame of G3 has a typical
mechanical precision of $\la 1$~mm over the length of 25 cm of the grid
support frame. This corresponds to an angular precision of $\sim 0.3^\circ$ which was confirmed by a direct measurement with an electronic
angle gauge in contact with the grid's frame.

Equation (\ref{def_angle}) shows that the precision of the $\tau$
measurement further depends on the correct {\it relative} calibration of
Stokes $U$ and $Q$. Since XPOL measures $Q$ as the power difference between
the H and V receivers, but derives Stokes $U$ from their correlation,
we need to make a precise measurement of the inevitable correlation
losses. This is done as a by-product of the frequent phase
calibration measurements where a wire grid of precisely known
orientation is observed against hot and cold loads. Due
to the high signal-to-noise of these calibrations, the correlation losses
can be determined with an \emph{r.m.s.} error of $\sim1$\%, resulting in an \emph{r.m.s.} error
of $\tau$ of $\sim0.3$$^\circ$.

The transfer of $\tau$ to the polarization angle $\alpha_{\rm Sky}$ as 
measured from north to east in the equatorial system is made through

\begin{equation}
\alpha_{\rm Sky} = 90^\circ + \tau - (\epsilon - \eta),
\label{nasmyth_to_equatorial}
\end{equation}

\noindent where the elevation $\epsilon$ and the parallactic rotation angle
$\eta$ do not introduce any additional error. We therefore estimate
that our polarization angles have an absolute precision on the order
of 0.5$^\circ$ or slightly better.


\section{Observations and data reduction}\label{obs}

\subsection{Mapping strategy}
The Stokes maps of the Crab nebula were taken in the on-the-fly mode
(hereafter OTF), allowing
us to map a large area on the sky within a reasonable time interval. The 
telescope was scanning the source while XPOL dumped twice per second the total
power from the orthogonally polarized receivers, and the complex
cross-correlation. The Stokes maps were subsequently derived from these
quantities. Throughout the paper, $Q$ and $U$ will
refer to the quantities measured in the Nasmyth system and then
rotated in equatorial coordinates according to equation \ref{nasmyth_to_equatorial}. The Crab nebula was scanned along the projections of latitude 
and longitude onto the local tangential plane (hereafter respectively
$\alpha$ and $\delta$ scanning directions). Such a scanning strategy is indispensable if
artifacts due to the variations of the atmospheric emission, leading to stripes 
in the maps, are to be removed (see section 4.2). The maps scanned along
the $\alpha$ direction are $8.8' \times 6.4'$ large, those scanned along
the $\delta$ direction $10.2' \times 5.5'$. The differences in size
are due to
slightly different system overheads coming from the acceleration and deceleration of 
the antenna mount drive before and after scanning an OTF stripe. 
The scanning speed was 16$''$/sec, leading to a smearing of the 27$''$ wide beam
(FWHM) by 8$''$. A pair of maps scanned in the orthogonal directions took
22 minutes, including regular temperature and phase calibrations. All
the maps were projected on the same
51$\times$41 pixels frame with square pixels of size 13.7$''$.

\subsection{Raw data reduction}
The conversion from backend count rates to antenna temperatures follows the
standard method applied at the IRAM 30 m observatory, involving total power measurements
of the sky, a hot load, and a cold load. These calibration measurements are
combined with the phase calibration measurements, namely an on-off between a
fully linearly polarized signal at cold load temperature, and the unpolarized
hot load signal at ambient temperature (\S~3). The calibrations were
derived and applied to the data with the MIRA software \citep[see][]{2009MIRA}.

Since the continuum emission from the Supernova remnant has a well-defined outer
boundary, there is no need to observe a separate emission-free reference
position. The atmosphere emission was rather removed by subtracting, from each
OTF subscan observed along a given direction, a linear baseline defined by the
Crab nebula emission-free map boundary. Subtracting only a zero order spatial baseline
would leave ``stripes'' in the maps along the scanning direction, which could
be removed in Fourier space with the PLAIT algorithm \citep{1988A&A...190..353E}.
Both methods were shown to yield equivalent results. In fact, they are
both limited by the uncertainty of the absolute reference flux if the source
boundary is not precisely known. In the image plane approach, the source and sky emission cannot be clearly separated from each
other, in the Fourier plane technique case, the PLAIT algorithm does
not work close to zero spatial frequency (whose Fourier component is the
total flux density contained in the map). 

  \begin{figure*}[t!]
    \centering
    \includegraphics[height=6cm,keepaspectratio]{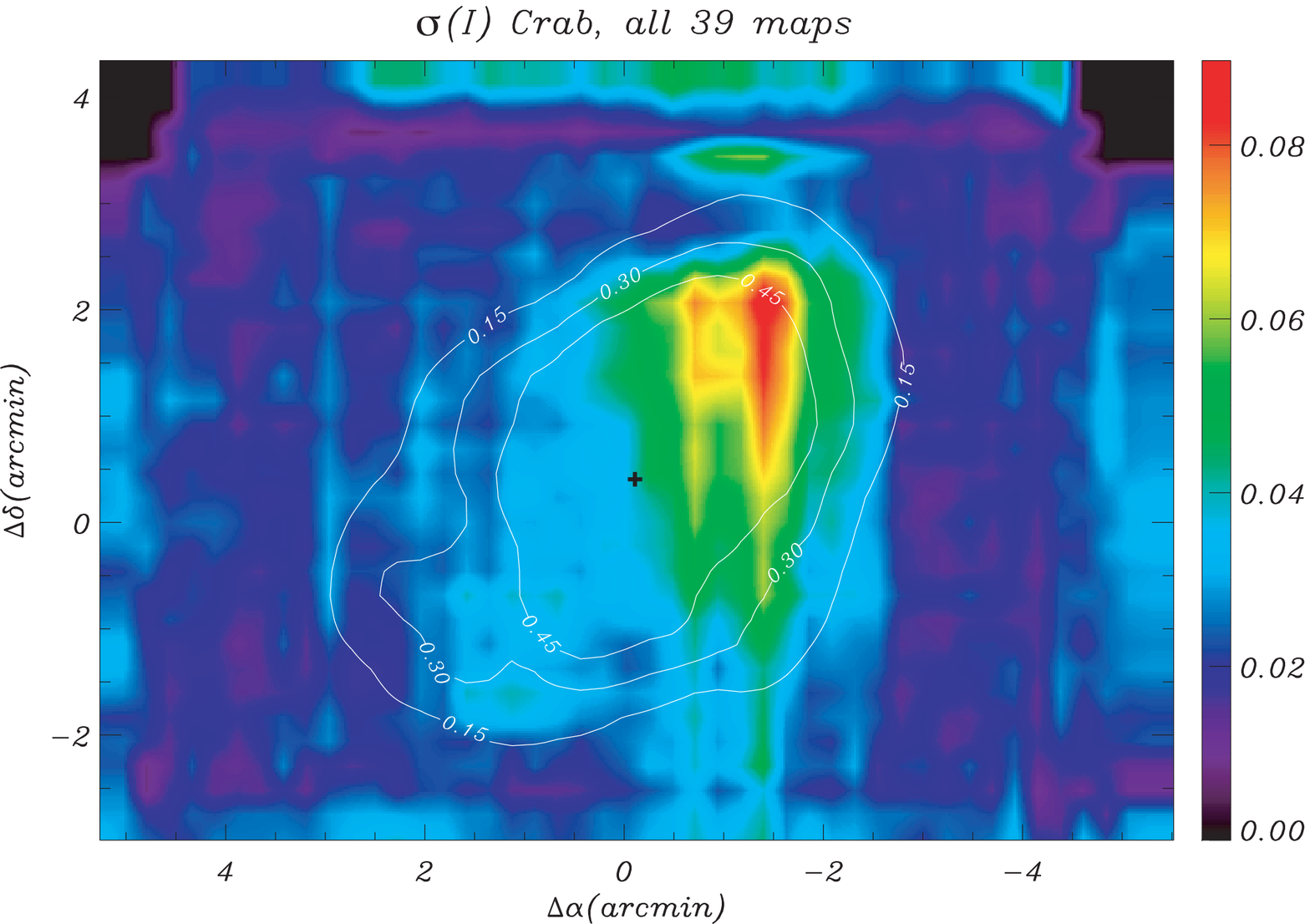}
    \includegraphics[height=6cm,keepaspectratio]{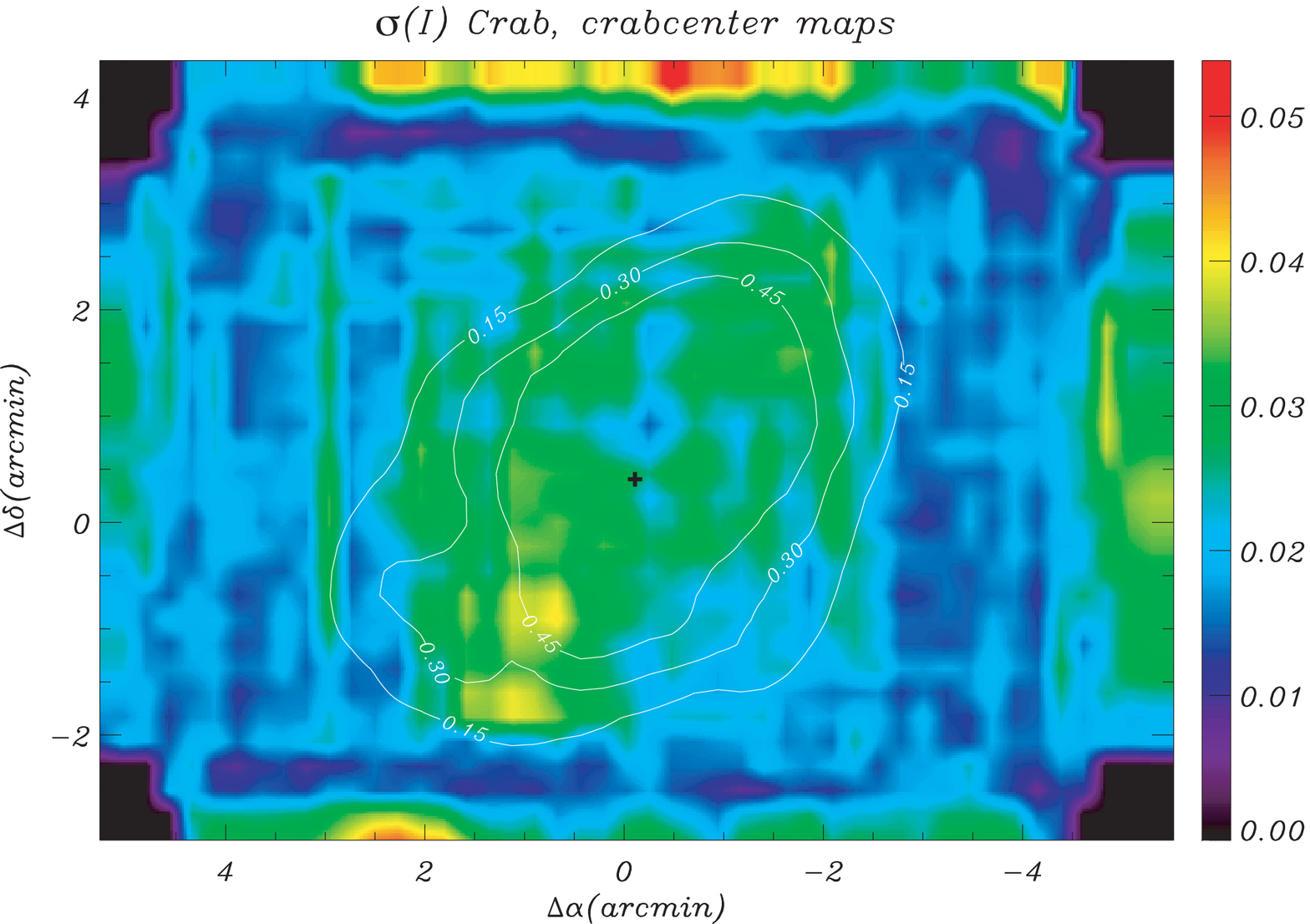}
    \includegraphics[height=6cm,keepaspectratio]{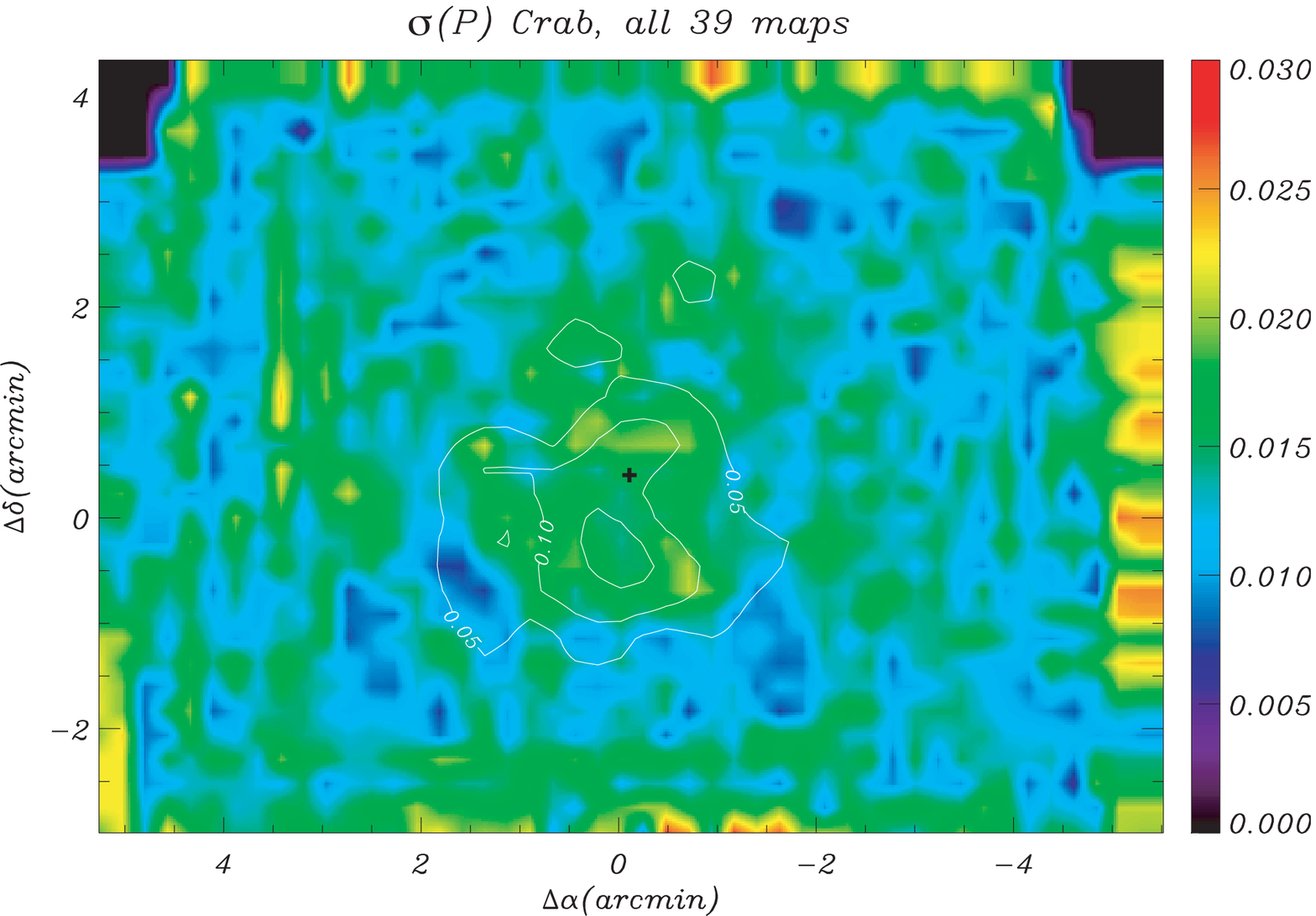}
    \includegraphics[height=6cm,keepaspectratio]{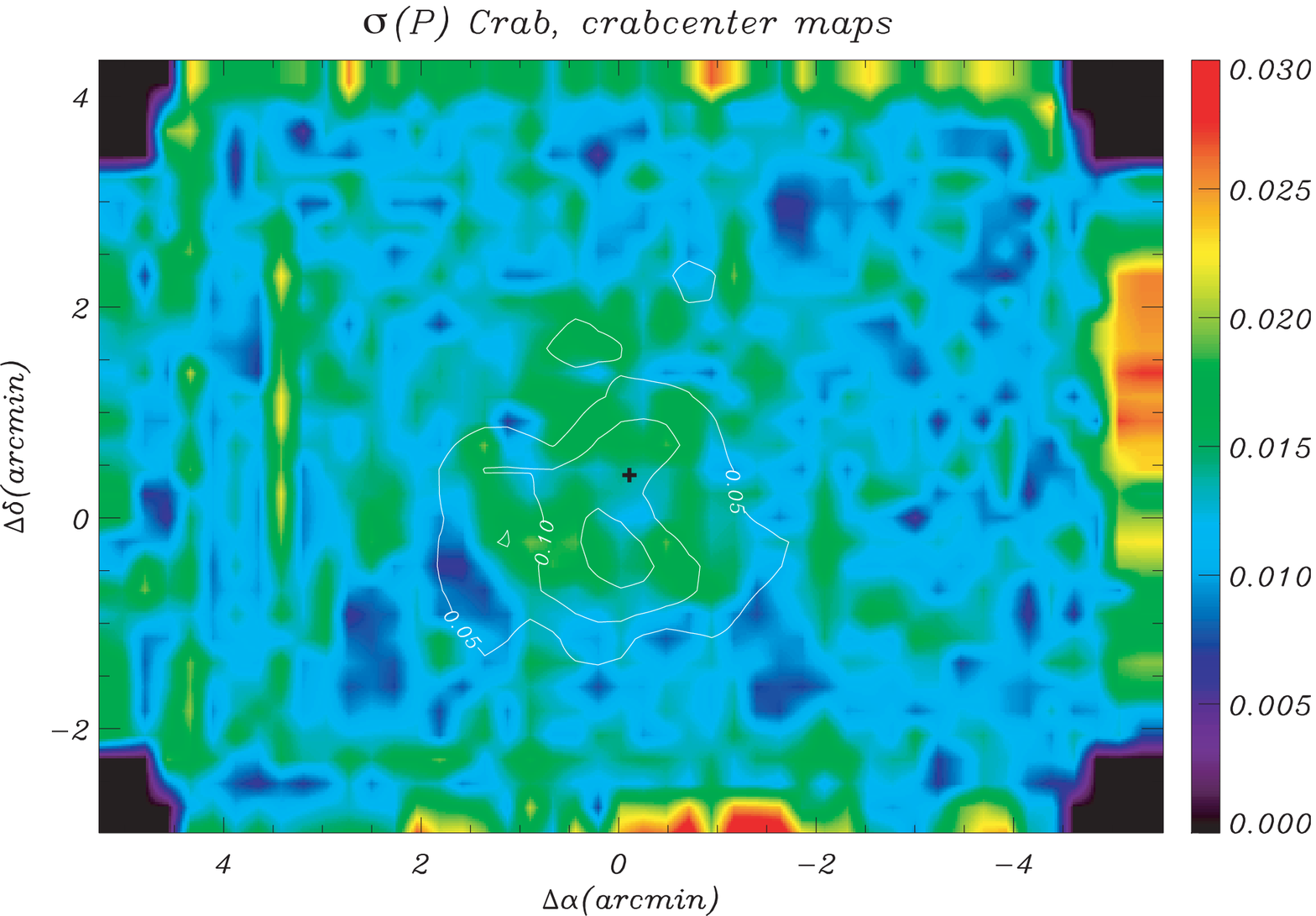}
    \caption{Standard deviation maps for the
      Crab nebula intensity $\sigma(I)$ (top row) and polarized intensity
      $\sigma(P)$ (bottom row), in antenna temperature (K). $\sigma(I)$ and $\sigma(P)$ maps were computed using the
      \emph{2CC} and \emph{2CX}
      (see text)
      39 maps (left column) and using the 28
      \emph{2CC} maps only (right panel). $I$ and respectively $P$ contours computed with the 28
      \emph{2CC} maps are overploted. We can see that when maps
scanned along the $\alpha$ direction are added to the data set the
standard deviation is higher and shows some features indicating an
uncontrolled systematic effect.}\label{fig:sip_all_ccent} 
  \end{figure*}

Since for XPOL the signal is downconverted into eight adjacent spectral
basebands of 62.5~MHz width each, total power variations between the
sky/hot/cold scan and the OTF record result in calibration mismatches among the 
basebands that introduce platforms in the autocorrelation spectra. In an attempt
to correct for this undesired effect, one may naively think of using the data
from a filterbank, connected in parallel and with a monolithic spectral band,
such that they are not affected by platforming. However, in the subsequent data 
processing step (rotation from the Nasmyth to equatorial reference
frame, for getting a stationary astronomical polarization), the spectra will be 
linearly combined with the real part of XPOL's cross correlation (which is,
after phase calibration, Stokes $U$ in the telescope's Nasmyth reference frame).
The small calibration differences observed between XPOL and the filterbank are
believed to be caused by non-linearities when the backends operate close to
the saturation limit. Mixing these data would thus lead to an inconsistent
calibration and hence an error in the polarization angle map of the Crab nebula.
Therefore, another strategy was used:

\begin{itemize}
\item For each autocorrelation spectrum (i.e. each record of an OTF map), the
median across the whole band, with concatenated basebands, has been calculated.
\item For each of the eight basebands of a given spectrum, the mean continuum
flux and noise has been determined.
\item When the absolute difference between the mean baseband signal and the
median signal is greater than twice the baseband noise, the offset of the baseband signal
has been corrected by the difference between the median and the baseband mean.
\end{itemize}

The result is a spectrum that looks ``flat''. The remaining calibration
mismatches lead to stripes in the maps, as do fluctuations of atmospheric
origin, and gain fluctuations owing to receiver instabilities. As a matter of
fact, stripes are also present in filterbank spectra, which do not suffer from
platforming.

It should be noted here that the cross-correlation spectra (Stokes $U$ in the
Nasmyth reference frame, and Stokes $V$) do not suffer from platforming either.
This derives from the fact that the clipping voltage of a sampler is adjusted
with respect to the signal's noise level, which may change when a given
interval of the time series is measured. The zero time lag channel
measures the total power (Parseval's theorem), which is positive
definite. After the FFT, this shows as an offset to the spectral baseline. A
cross-correlator analyses the signals from two independent samplers.
Variations in the noise power of the signals coming from the samplers are
largely uncorrelated, and therefore cancel out in the spectral cross-power
(which is not a positive definite quantity anymore).

The data were gridded at Nyquist sampling. Because of the slightly different
values of the start or end scanning coordinate between different subscans, the
sampling is irregular. Therefore the data had to be resampled to a regular grid. This
has been done by applying a Gaussian convolution kernel with a FWHM of one-third
the telescope's half-power beam width. The kernel is truncated at
three FWHM.

\subsection{Description of the data sets}

We performed two observation campaigns of the Crab Nebula,
respectively from September 5 to 10,
2007, and from January 9 to 12,
2009.

\begin{itemize}
\item \emph{First campaign}: Observations were made during 6 intervals,
under varying astmospheric opacity and stability conditions. Most of
the maps were affected by a substantial atmospheric contamination,
showing as strong linear stripes in the scanning direction. We
  obtained 66
  $I$, $Q$, $U$ and $V$ maps resulting in $8' \times
6.4'$ maps after coadding $\alpha$-scanned and $\delta$-scanned maps. These maps were centered on the reference point we called
\emph{crabxpol} having equatorial
coordinates $\alpha = 5^{\rm h} 34^{\rm m} 31.5^{\rm s}$ and $\delta =
22^\circ 00\arcmin 27.7\arcsec$ (J2000). The centering and the size of these
maps made the destriping delicate because the extension of the source emission
was larger than the observed area and thus, strong mix-up with
atmospheric contamination arose on maps' edges leading to uncontrolled
systematics.

\item \emph{Second campaign}: Observations were made during 4 nights
  during which we experienced stable atmospheric
  conditions. As a consequence of the first observation campaign, we increased the coaddition of the $\alpha$-scanned and
  $\delta$-scanned maps to a size of $10' \times
6.7'$ in order to be able to
  characterize the atmospheric stripes outside the source. 11 maps
  of $I$, $Q$, $U$ and $V$ were observed centered on the same
  reference point as in the first campaign, namely \emph{crabxpol},
  and then we changed to a new reference point for center, called
  \emph{crabcenter} and having equatorial coordinates $\alpha = 5^{\rm
h} 34^{\rm m} 31.5^{\rm s}$ and $\delta = 22^\circ 01\arcmin
17.7\arcsec$ (J2000), on which the next 28 maps were
observed. This choice was made to have a better centering of the
source extent in the north-south direction, allowing better
destriping. The 11 first maps centered on \emph{crabxpol} have a
significantly higher noise level than the 28 centered on
\emph{crabcenter} due to worse weather condition at the beginning of
the campaign.

\end{itemize}

These data sets thus gather 105 individual $I$, $Q$, $U$ and $V$
maps, divided in three subsets: first campaign maps centered on
\emph{crabxpol} (hereafter \emph{1CX}), second campaign maps centered on
\emph{crabxpol} (hereafter \emph{2CX}) and second campaign maps centered on
\emph{crabcenter} (hereafter \emph{2CC}). Because in the two first
subsets the source extends over an area larger than the mapped one,
the destriping is delicate and thus may lead to uncontrolled
systematics. For each subsets and for combinations of different subsets we computed, for each Stokes parameter, mean and
dispersion for each pixel among the whole data set. As the
pixels are small with respect to the beam, the computed dispersion does not
suffer from bias. Dispersion maps,
in the case of the first and second subsets, present specific
high-dispersion features
arising from the confusion between stripes and source emission on the
borders during the destriping procedure. This can be seen on Figure
\ref{fig:sip_all_ccent} where the combination of subsets \emph{2CX} and \emph{2CC}, for
the intensity, shows a higher standard deviation feature than the
\emph{2CC} alone. For consistency, we
also checked the behavior of the polarized intensity standard
deviation which slightly decreases when using only
\emph{2CC} maps.

Thus, using only the 28 maps from
\emph{2CC} provides the best results and any addition
of \emph{2CX} maps, only
\emph{2CX} maps scanned in $\alpha$ direction
(direction in which the borders of the maps are off-source) and
\emph{1CX} lead to worse results in terms of noise
homogeneity and of signal-to-noise ratio, both in intensity and
polarization. Consequently, our analysis will use only the 28
maps of the second campaign centered on the \emph{crabcenter}
reference point.

\subsection{Systematic effects analysis}
\label{syste_analysis}

In order to check for residual systematic effects in the maps, we
performed jack-knife tests dividing the 28 maps in several subsets using three
criteria, i.e.: randomly, between maps scanned in $\alpha$ and $\delta$
direction and between the three observation nights during which these
28 maps were taken. We also divided the data in two with respect
to the position angle of the Nasmyth reference frame projected onto
the sky, $\chi_0$, in order to check for sidelobe polarization
effects \citep{2008A&A...492..757F}, which may show whether the difference between the $\chi_0$ of
subsets is close to 90$^\circ$.

\begin{table}[t]
        \centering
        \begin{tabular}{l|c|c}
        Data subset            & Pol. angle $\alpha_{\rm Sky}$
        & Pol. fraction $\Pi$ \\
        \hline
        \hline
        Random 1                & $152.19 \pm 0.36^{\circ}$     & $23.77 \pm 0.52$\% \\
        Random 2                & $152.07 \pm 0.30^{\circ}$     & $23.91 \pm 0.54$\% \\
        \hline
        $\alpha$ scans & $152.02 \pm 0.32^{\circ}$     & $23.38 \pm 0.45$\% \\
        $\delta$ scans           & $152.26 \pm 0.35^{\circ}$     & $24.41 \pm 0.53$\% \\
        \hline
        Night 1                   & $152.47 \pm 0.37^{\circ}$     & $23.96 \pm 0.55$\% \\
        Night 2                             & $151.89 \pm 0.31^{\circ}$     & $23.93 \pm 0.56$\% \\        
        Night 3                             & $152.10 \pm 0.30^{\circ}$     & $23.56 \pm 0.49$\% \\
        \hline
        $\chi_0\in\{-30^\circ,45^\circ \}$                            & $152.17 \pm 0.37^{\circ}$     & $23.60 \pm 0.52$\% \\
        $\chi_0\in\{45^\circ,120^\circ \}$                            & $152.09 \pm 0.30^{\circ}$     & $24.12 \pm 0.55$\%\\
        \hline
        Whole data set                   & $152.13 \pm 0.34^{\circ}$     & $23.84 \pm 0.54$\% \\
        \end{tabular}
        \caption{Polarization angle $\alpha_{\rm Sky}$ and polarized
          fraction $\Pi$ computed for different divisions of the data set.}\label{table:systematics}
\end{table}

\begin{figure*}
  \centering
  \includegraphics[height=6cm,keepaspectratio]{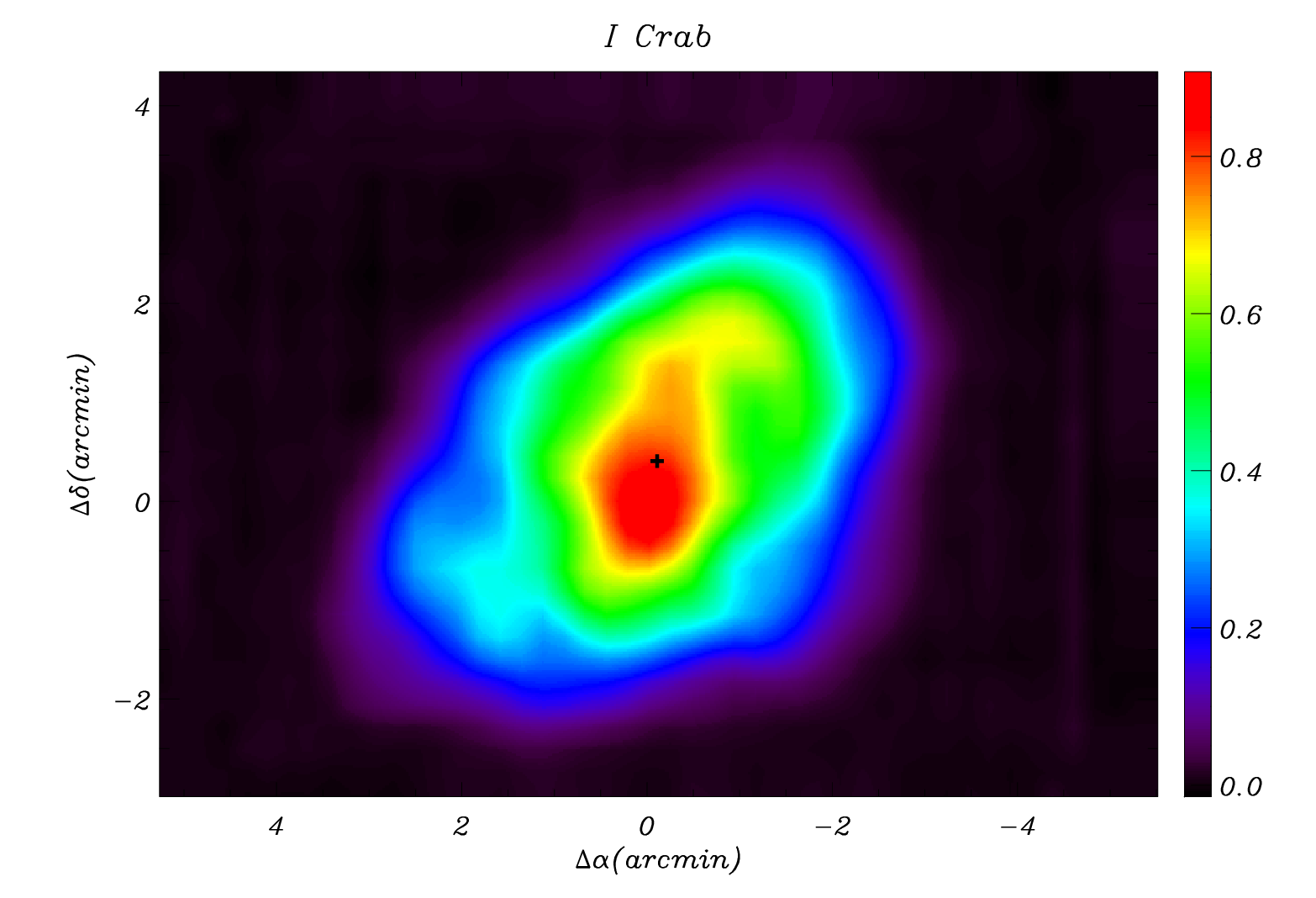}
  \includegraphics[height=6cm,keepaspectratio]{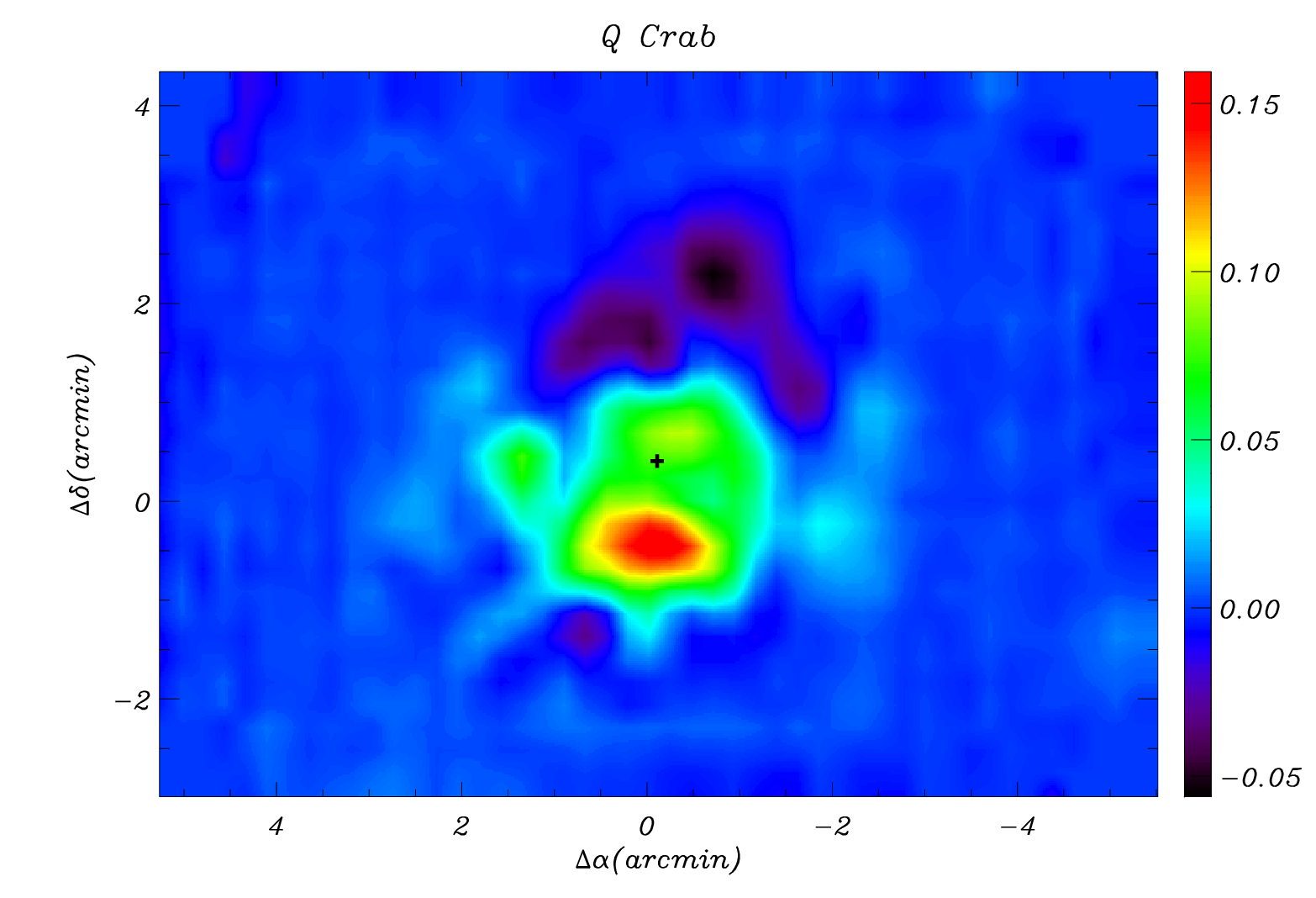}
  \includegraphics[height=6cm,keepaspectratio]{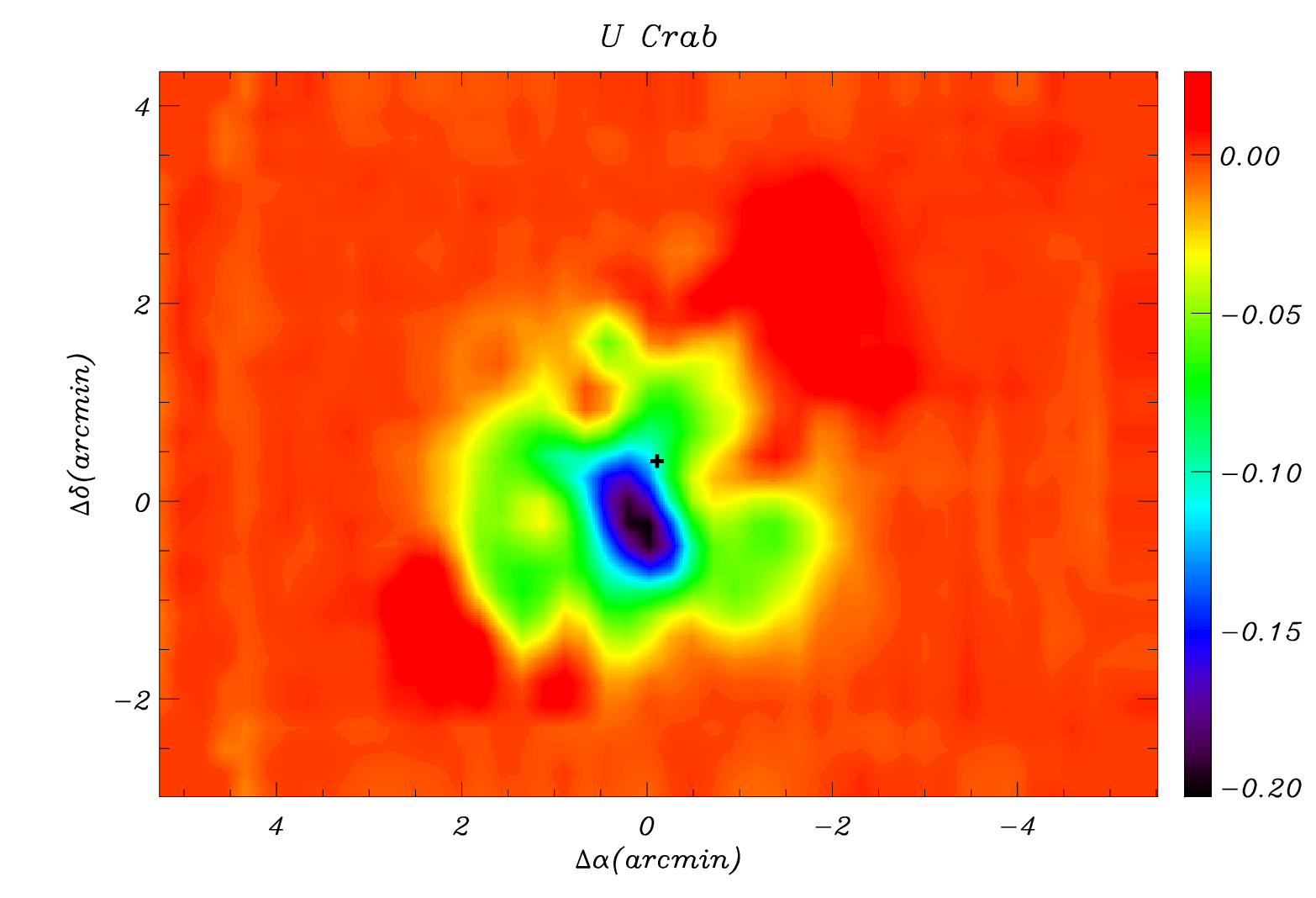}
  \includegraphics[height=6cm,keepaspectratio]{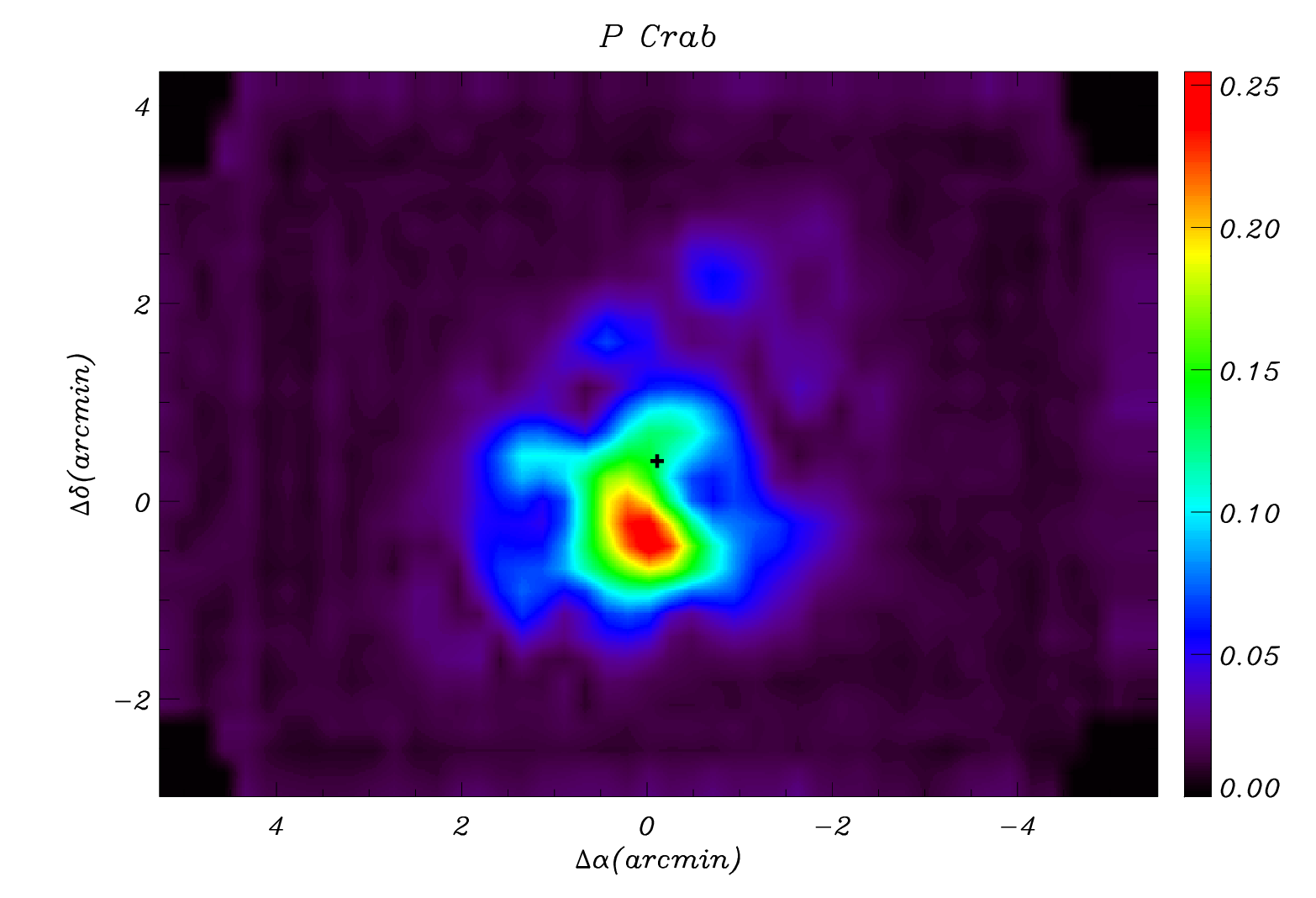}
  \caption{Maps of the Crab nebula at 89.189 GHz in antenna temperature (K) for
    intensity $I$ (top-left), $Q$ polarization (top-right), $U$
    polarization (bottom-left) and polarized intensity $P$
    (bottom-right). The position of the Crab nebula pulsar is
    indicated by the black cross.}\label{fig:iqup_ccent} 
\end{figure*}

The mean and dispersion of the 28 observations allow us to derive a signal-to-noise map for $I$, $Q$, $U$ and
$P\equiv(Q^2+U^2)^{1/2}$. For the residual systematic effects
analysis, we have chosen a region corresponding to S/N($P$)$>10$ that
includes 19 pixels. For these pixels, we compute the mean polarization angle
$\alpha_{{\rm Sky},j}\equiv0.5\cdot\langle\arctan(U_j/Q_j)\rangle$ and the mean
polarization fraction $\Pi_j\equiv\langle P_j/I_j\rangle$ for each
map $j$. Then, $\alpha_{{\rm Sky},j}$ and $\Pi_j$ are averaged for each
subset to give the mean $\alpha_{\rm Sky}$ and $\Pi$ and the error on
the mean is given by the standard deviation over the subset of
maps, preventing correlations between $Q$ and $U$, or among pixels.

Results are displayed in Table \ref{table:systematics}. Values are
stable against the selection of different data
subsets. For the polarization angle, for each subset selection criteria, the values are compatible at less than two $\sigma$, showing no clue to residual
systematic effects in the data. We observe a similar behavior for the
polarization fraction, where only the subset corresponding to
$\alpha$ and $\delta$ scans shows a departure of slightly more than two $\sigma$. This systematic behavior is due to a
percent-level difference in the intensity maps corresponding to each
type of scan, caused by residual errors in the destriping which is less
precise for $\delta$-scanned maps due to the limited number of
off-source pixels in this case and leading to an overestimation of the intensity.

\section{Results}\label{results}

\begin{figure*}
  \centering
  \includegraphics[height=6cm,keepaspectratio]{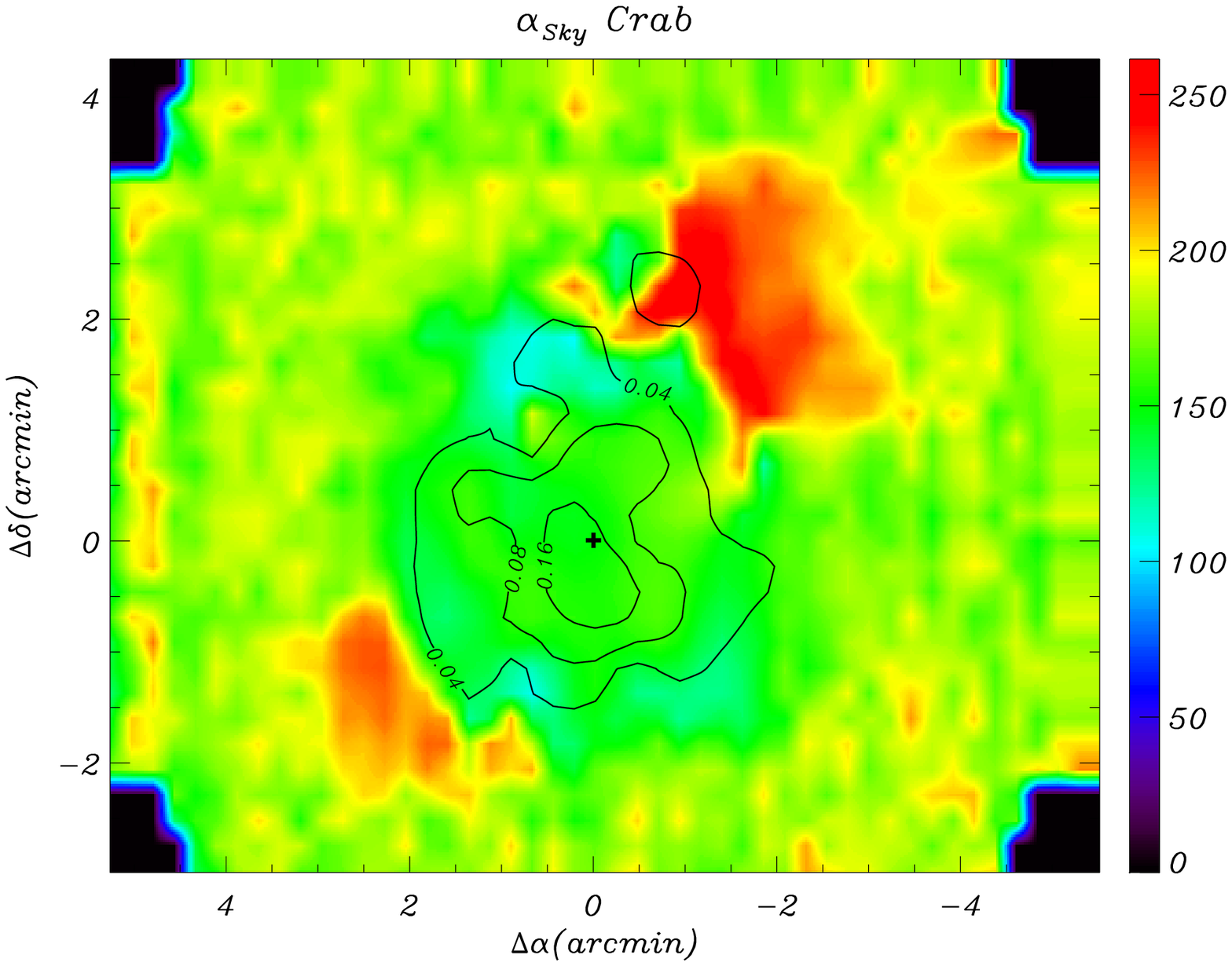}
  \includegraphics[height=6cm,keepaspectratio]{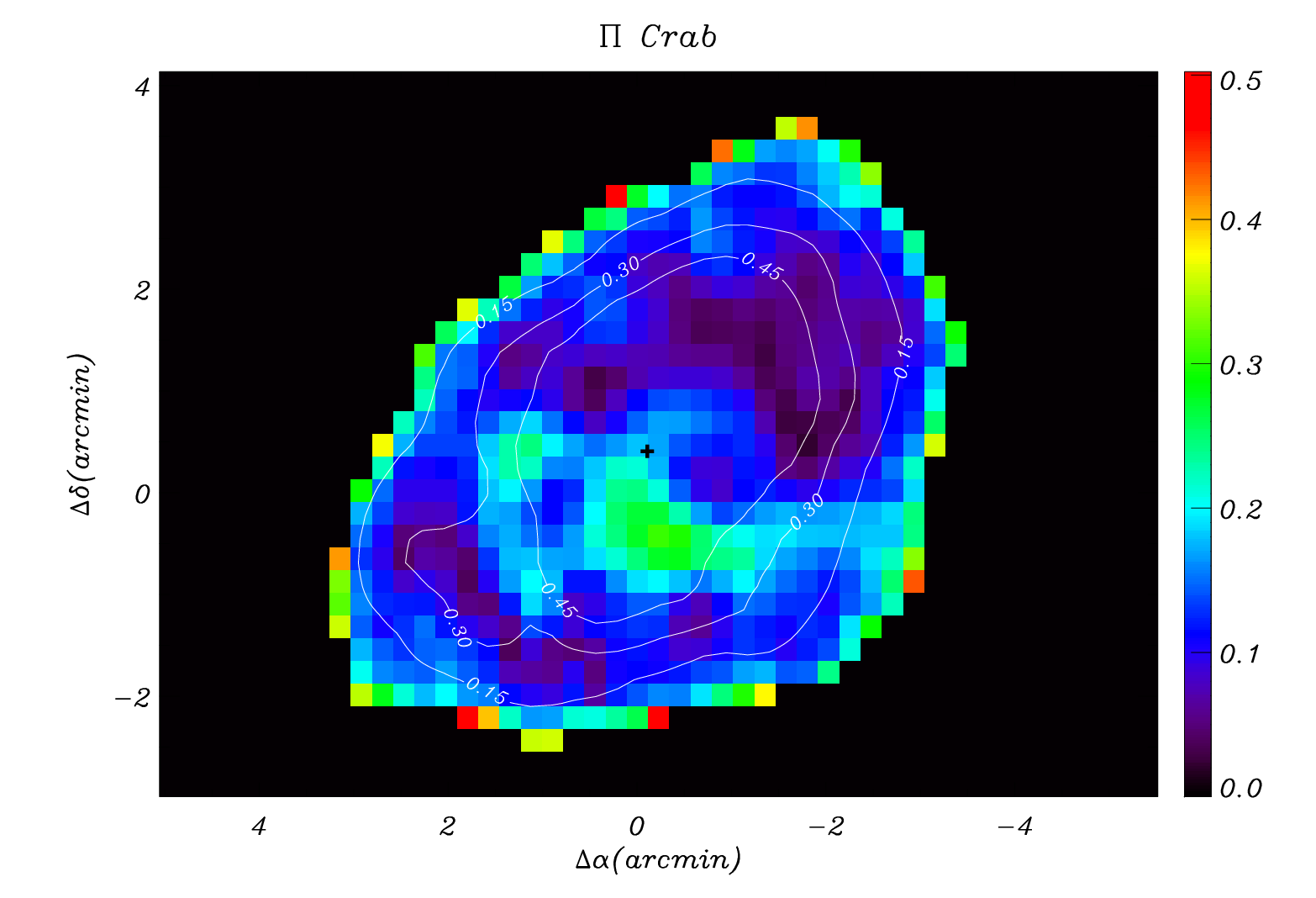}
  \caption{(left): Map of the polarization angle $\alpha_{\rm Sky}$ in degree on which contours of the polarized intensity $P$ for 0.04,
    0.08 and 0.16 K have been
    overploted. (right): Map of the polarization fraction $\Pi$ on which
  has been displayed the intensity contour for 0.15, 0.30 and 0.45 K.}\label{fig:gpi_ccent} 
\end{figure*}

After the raw data reduction, including atmospheric decontamination by
linear baselines subtraction, we obtained a set of 28
systematic-reduced individual maps with which we compute average and standard
deviation maps for $I$, $Q$, $U$, $P\equiv(Q^2+U^2)^{1/2}$,
$\alpha_{\rm Sky}\equiv0.5\cdot\arctan(U/Q)$ and $\Pi\equiv P/I$. 

The Crab nebula intensity $I$ map is displayed on top-left panel of Figure
\ref{fig:iqup_ccent}, in antenna temperature, showing a maximum 
0.91 K emission half an arcmin south from the pulsar position. The
flux density integrated over the source and its error bar derived from
the pixel-to-pixel standard deviation are
measured as $S(I)=195.5\pm11.0$ Jy. This value is compatible within
3$\sigma$ with the
value given by the WMAP satellite at 92.9 GHz of $S(I)=229\pm11$ Jy
\citep{2007ApJS..170..335P}.
 
The $Q$ and $U$ maps of the Crab nebula are displayed on top-right and
bottom-left panel of Figure \ref{fig:iqup_ccent}, in antenna temperature
(K). $Q$ map is
showing a roughly positive value, up to 0.16 K while $U$ is showing
 a negative one down to -0.20 K.

The polarized intensity $P$ map is displayed on bottom-right panel of Figure
\ref{fig:iqup_ccent}, in antenna temperature, showing a maximum of
0.25 K half an arcmin south from the pulsar position. We can see that
the Crab nebula is less extended in polarized intensity than in Stokes
$I$. The polarized flux density is
measured as $S(P)=14.5\pm3.2$ Jy, leading to a mean polarization
fraction for the whole Crab nebula source of $\Pi=7.4\pm0.7$ \%. This
value is compatible with the value measured by WMAP of $7.6\pm2.0$ \% \citep{2007ApJS..170..335P}.

\begin{table}[h]
        \centering
        \begin{tabular}{c||c|c}
          & $\Pi$ (\%) & $\alpha_{\rm Sky}$ ($^\circ$)\\ 
          \hline
          Pulsar position &$13.9\pm0.6$ & $158.1\pm0.5$\\
          S/N($P$)$>10$ region &$23.8\pm0.5$ &$152.1 \pm0.3$\\  
          S/N($P$)$>3$ region &$15.6\pm0.3$ &$153.7\pm0.4$\\
          seen by a 5' beam &$8.8\pm0.2$ &$149.9\pm0.2$\\
          seen by a 10' beam&    $7.7\pm0.2$&$148.8\pm0.2$\\
        \end{tabular}
        \caption{Values of the polarization angle $\alpha_{\rm Sky}$ and
          of the polarization fraction $\Pi$.}\label{tab:sky_values}
\end{table}

The $\alpha_{\rm Sky}$ map is displayed on the left panel of Figure
\ref{fig:gpi_ccent}. It is worth noting that $\alpha_{\rm Sky}$ is
almost constant
around 150$^\circ$ in the region of maximum polarized intensity
($P>0.05$ K). This region corresponds to the region where each pixel's standard
deviation $\sigma(\alpha_{\rm Sky})$ is the lowest ($\le30^\circ$), going
down to $\sigma(\alpha_{\rm Sky})\lesssim3^\circ$ for the most intensely
polarized pixels. The value of the angle we measured at the flux peak,
$\alpha_{\rm Sky}=149.0\pm1.4^\circ$, is in very good agreement with previous measurements at 1350 and
1100 $\mu$m \citep{1991MNRAS.249P...4F,2003MNRAS.340..353G}. On the map edges,
$\alpha_{\rm Sky}$ pixel-to-pixel deviation is high, except for the
northern region where the average angle is around 230$^\circ$. When
looking at the $\sigma(\alpha_{\rm Sky})$ map, we can see that pixels
on the edges have an undetermined value ($\sigma(\alpha_{\rm
Sky})\simeq90^\circ$), whereas in the northern
region they have $30^\circ\le\sigma(\alpha_{\rm Sky})\le50^\circ$,
indicating a polarization angle there that is significantly
different from that near the center of the nebula.

The polarization fraction $\Pi$ map is displayed on the right panel of Figure
\ref{fig:gpi_ccent}. We have set to 0 the pixels for which intensity
is lower than 0.02 K in order to avoid the divergence of $\Pi$. Maximum
polarization fraction is found to be spatially correlated with maximum
polarized intensity region, with a polarization fraction reaching 30\% for few
pixels. A 1.5 arcmin circular region around the maximum polarization fraction has a significant $\Pi\sim20$\%.

\begin{figure*}
  \centering
  \includegraphics[height=9cm,keepaspectratio]{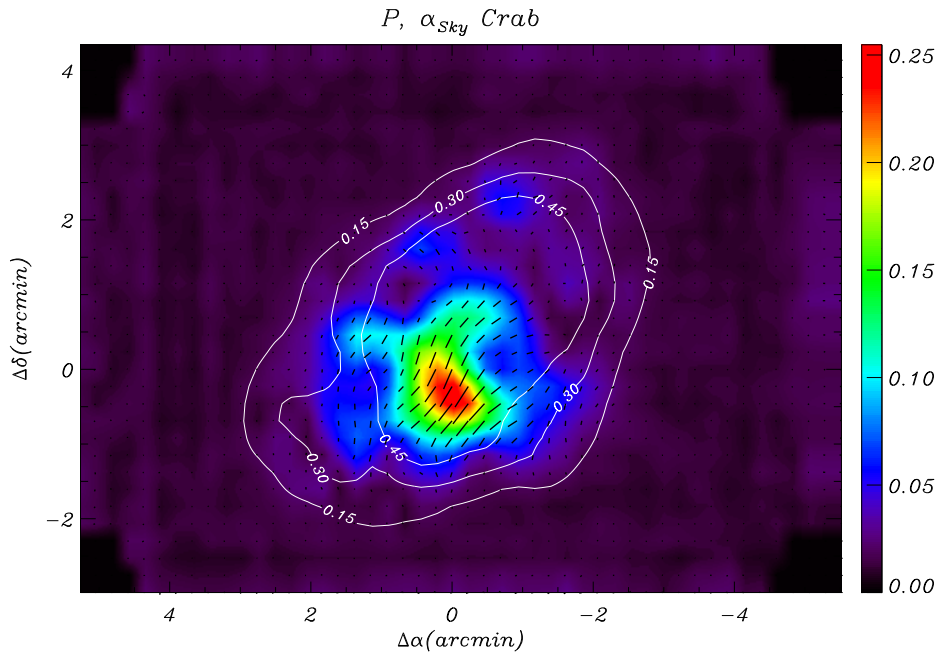}
  \caption{Map of the Crab nebula polarized intensity $P$ at 89.189
    GHz in antenna temperature on which polarization vectors have been
  overpolted. Intensity contours at 0.15, 0.30 and 0.45 K are also displayed.}\label{fig:pgi_ccent} 
\end{figure*}

To summarize all these results, we display in Figure
\ref{fig:pgi_ccent} the polarized flux density $P$ and the orientation
of the polarization vectors associated with each map pixel. The
polarization vectors pattern is compatible with 9 mm \citep{1979MNRAS.189..867F} and 850
$\mu$m \citep{2003MNRAS.340..353G} observations.

We finally
computed several values of the polarization angle $\alpha_{\rm Sky}$ and
of the polarization fraction $\Pi$ for different regions, displayed in
Table \ref{tab:sky_values}. We selected the pulsar position and high signal-to-noise ratio regions for
the polarized intensity (S/N($P$)$>3$, corresponding roughly to the
source's polarized emission extent and S/N($P$)$>10$). Furthermore,
we convolved
each individual map by a circular Gaussian beams of 5 arcmin and 10
arcmin of FWHM before computing the mean polarization angle and fraction. These
beams mimic the Planck satellite's beams (10$'$ for the 100 GHz and 5$'$
for the 217 and 353 GHz channels) and give an estimation on how a
generic CMB experiment having Planck-like characteristic would see the
Crab nebula's polarization. For each of these cases, the mean value
and its associated error were computed similarly to section
\ref{syste_analysis}.


\section{Extension to other frequencies}\label{other_freq}

The total intensity emission of the Crab nebula from 1 to $10^6$
GHz is dominated by the
well known synchrotron radiation observed at radio wavelengths with
only one extra dust component in the far infrared  \citep{2009MaciasPerez}. This synchrotron emission shows a  
decrease of flux with increasing frequency which can be represented 
by a power law of spectral index $\beta = -0.296 \pm 0.06$
\citep{Baars1977,2009MaciasPerez} from the radio to the submillimeter domains. The flux
is also decreasing with time at a rate of $\alpha = 0.167 \pm 0.015$
\% yr$^{-1}$ \citep{Aller1985}. Moreover, from the visible to the X-rays the
synchrotron emission evolves towards a much harder spectrum
represented by a power law of spectral index $\beta = -0.698 \pm 0.018$
\citep{2009MaciasPerez}.

These statements allow us to postulate that where the synchrotron
dominates, the emission at different wavelengths is produced by
particles accelerated by the same magnetic
field. The direction of polarization is thus expected to be constant
while the polarization fraction may vary. Nevertheless, we expect
values for both the polarization angle and the polarization fraction,
at low-resolution, to be
similar in the millimeter and submillimeter where the emission is
produced by the same electron population.

Values of the polarization angle of the Crab nebula source have been reported over this wide
range of wavelengths from the radio \citep[e.g. at 9 mm, $\alpha_{\rm
Sky}=154.8\pm2.0$, ][]{1979MNRAS.189..867F} to the millimeter (at 3.3 mm, $\alpha_{\rm
Sky}=153.7\pm0.4$, \emph{this paper}) and to the X-rays \citep[at 240 pm, $\alpha_{\rm
Sky}=155.8\pm1.4$, ][]{1978ApJ...220L.117W}. We can see that the value
of the polarization angle is strikingly constant over nearly ten
decades of wavelength. Furthermore, the value measured in the X-ray,
similar to the one we measured at 90 GHz, indicates that we are
probably dominated by synchrotron emission for the same regions in
both cases and that an extrapolation up to 353 GHz, where
present and future CMB experiments such as the Planck satellite are
observing the sky, is rather safe. 

The polarization fraction can be compared to other measurements too,
keeping in mind that it is a quantity which changes inside the source
rapidly and that shall be compared only for
experiments observing with a comparable beam size. The
comparison of our value of $\Pi=15.6\pm0.2$\% to radio observations
at 9 mm giving $\Pi=16\pm1$\% \citep{1979MNRAS.189..867F} and the value we obtain when our maps were convolved by a 10$'$
beam of $7.7\pm0.2$\% to the WMAP experiment \citep[$7.6\pm2.0$ \%,
][]{2007ApJS..170..335P} indicates that those measurements are
coherent with an emission coming from the same electron population
wich leads to a constant polarization fraction over the CMB frequency
range.


\section{Summary}\label{conclusions}

We mapped the polarized emission of the Crab nebula using the IRAM 30m telescope at 89.189 GHz
with an angular resolution of 27$''$, with two orthogonally linearly
polarized heterodyne receivers. The Stokes parameters were derived
from the auto- and cross-correlations using the XPOL procedure
\citep{2008PASP..120..777T}. 

Observations have been undertaken during two campains, leading to a
set of 105 individual $I$, $Q$ and $U$ maps having changing weather
conditions, sizes and centering. Particular care has been taken in
choosing the set of individual maps leading to the lower level of
systematics in the final products. Additionnaly, jack-knife tests have
been carried out in order to demonstrate the robustness of the
data. As a result, we constructed 10$'\times6.7'$ $I$, $Q$ and $U$ coaddition maps of these
observations centered on the Crab nebula.

From these maps we have computed the polarized intensity
$P=\sqrt{Q^2+U^2}$, the polarization angle $\alpha_{\rm
Sky}\equiv0.5\cdot\langle\arctan(U/Q)\rangle$ and the polarization
fraction $\Pi\equiv\langle P/I\rangle$. We derived our results on the
polarization characteristics of the Crab nebula with these maps:

\begin{itemize}

\item The measured flux density is $S(I)=195\pm11$ Jy
and the polarized flux density is $S(P)=14.5\pm3.2$ Jy.
\item The polarization angle is almost constant in the region of
maximum emission in polarization with a mean value of $\alpha_{\rm Sky}=152.1\pm 0.3^\circ$. A region north to the pulsar is seen with an average angle of
$\sim$230°, but it does not correspond to a strongly polarized emission
region. When seen by a 5 arcmin beam, the mean polarization angle of
the whole source as a value of $\alpha_{\rm Sky}=149.9\pm 0.2^\circ$,
in good agreement with the other measurements at radio, millimeter and
X-rays wavelengths. 
\item The polarization fraction shows a maximum in a region south from
the pulsar position, reaching 30\%. When averaged by a 5 arcmin beam,
the measured mean value of the polarization fraction is
$\Pi=8.8\pm0.2$ \%. This value is in very good agreement with the
WMAP measurements at 94 GHz. 

\end{itemize}

With a precision of a few tenths of degree on the polarization angle
and of a few tenths of percent on the polarization fraction,
these measurements will be especially interesting in the purpose of absolute
calibration of the polarization angle and the cross-polarization
leakage for present and future polarized CMB experiments, such as the Planck mission, in order to accurately disentangle instrumental and
systematic $B$-modes polarization from the primordial one.

\begin{acknowledgements}
We gratefully acknowledge the support of IRAM to this program
including the grant of discretion time by its director P.
Cox. J.Aumont has been partly supported in this work by a
post-doctoral position from CNES. E. Pointecouteau was supported by grant
ANR-06-JCJC-01142.

\end{acknowledgements}

\bibliographystyle{aa}
\bibliography{13834}

\end{document}